\documentstyle[12pt,aasms4]{article}

\def\ltsima{$\; \buildrel < \over \sim \;$}
\def\simlt{\lower.5ex\hbox{\ltsima}}
\def\gtsima{$\; \buildrel > \over \sim \;$}
\def\simgt{\lower.5ex\hbox{\gtsima}}
 
\received{}
\accepted{}
\journalid{}{}
\articleid{}{}
\slugcomment{To appear in ApJ}
\lefthead{}
\righthead{}

\begin{document}
   
\title{How Abundant is Iron in the Core of the Perseus Cluster?}   
\author{S.Molendi\altaffilmark{1,2}, G.Matt\altaffilmark{3},         
        L.A.Antonelli\altaffilmark{2}, F.Fiore\altaffilmark{2,4}, 
        R. Fusco-Femiano\altaffilmark{5}, J.Kaastra\altaffilmark{6}
        C.Maccarone\altaffilmark{7} and C.Perola\altaffilmark{3}
        }

\altaffiltext{1}{Istituto di Fisica Cosmica, CNR, via Bassini 15,
I-20133 Milano, Italy}

\altaffiltext{2}{BeppoSAX Science Data Center, ASI,
    Via Corcolle, 19
    I-00131 Roma , Italy}

\altaffiltext{3}{Dipartimento di Fisica ``E.Amaldi'', Universit\`a degli Studi 
    ``Roma III, Via della Vasca Navale 84, I-00146 Roma, Italy}

\altaffiltext{4}{Osservatorio Astronomico di Roma Monteporzio Catone,
      Via Osservatorio, 2 I-00040 Monteporzio Catone (Roma),
      Italy}

\altaffiltext{5}{Istituto di Astrofisica Spaziale, CNR,
   Via Enrico Fermi, 21
    I-00044 Frascati RM, Italy}

\altaffiltext{6}{Space Research Organization, Sorbonnelaan, 
    2 NL-3584 CA Utrecht, The Netherlands}

\altaffiltext{7}{Istituto di Fisica Cosmica, C.N.R., Via La Malfa 153, I-90146  
   Palermo, Italy}

\begin{abstract}

The analysis of Perseus data collected with the Medium Energy 
Concentrator Spectrometer (MECS) on board Beppo-SAX shows that
the ratio of the flux of the 8 keV line complex (dominated by 
Fe K$_{\beta}$ emission) over the 6.8 keV line complex (dominated by 
Fe K$_{\alpha}$ emission) is significantly larger than predicted by standard 
thermal emission codes. Moreover the analysis of spatially resolved spectra
shows that the above ratio decreases with increasing cluster radius. 
We find that, amongst the various explanations we consider, the  most 
likely requires the plasma to be optically thick for resonant
scattering at the energy of the Fe K$_{\alpha}$ line.
We argue that if this is the case, then measures of the 
iron abundance made using standard thermal emission codes,
that assume optically thin emission, can significantly underestimate 
the true iron abundance. In the case of the core of Perseus we estimate 
the true abundance to be $\sim$ 0.9 solar in a circular region 
with radius of $\sim 60$ kpc and centered on NGC 1275. 
Finally we speculate that similar results may hold for the core of 
other rich clusters. 

\end{abstract}


\section {Introduction}
The X-ray emission from clusters is due to a diffuse thermal plasma 
permeating the intra-cluster space. The plasma
is tenuous, with typical densities of $ 10^{-4}$ - 10$^{-2}$ cm$^{-3}$,
hot, with temperatures in the range 10$^7$ - 10$^8$ K, and 
optically thin at almost all energies.  Under these conditions the 
plasma is so ineffective in radiating its thermal energy that it cools 
down on timescales comparable to the
age of the Universe. Thus it is quite plausible that a large fraction
of the emitting gas in clusters is of primordial origin, in the sense that
it has never been cycled through stars. Such a possibility is actually
corroborated by measurements of the heavy elements abundances, 
principally iron, in galaxy clusters.  Indeed, Fe abundances typically 
range between  0.3 solar and 0.4 solar. 
Iron abundance estimates are mostly derived from measurements of the 
equivalent width of the prominent Fe K$_{\alpha}$ line complex at $\sim$ 
6.8 keV. 
It can be easily show 
(e.g. Sarazin 1988) that under the assumption that the gas is optically 
thin, the equivalent width of an emission line is directly proportional 
to the abundance of the element responsible for the line emission. 

However, as pointed out by Gilfanov et al. (1986; hereinafter G86) 
emission at the Fe K$_{\alpha}$ line energy may not be always optically thin.
Resonant scattering, i.e. the process describing the absorption of an 
Fe K$_{\alpha}$ line photon by an iron ion followed by the 
immediate re-emission, can be quite effective for typical cluster gas 
densities and temperatures. The above authors have shown that
the cores of rich cluster, such as Perseus and Virgo, should have 
optical depths, for the resonant scattering process, of the order
of a few. 
Indeed, if the gas is optically thick to resonant scattering, 
the line emission coming from the core of the cluster will be attenuated
because of the photons which are scattered out of the line of sight. 
If this is the case then the abundances
measured using standard thermal emission codes, which
assume optically thin thermal emission, can be significantly 
underestimated.

The attenuation of the line intensity cannot be directly measured
using the Fe K$_{\alpha}$ line only, at least not with current 
instrumentation \footnote{ 
Resonant scattering produces a depression at the core of the line with a 
characteristic width of (G86)
$ \Delta \nu \sim \nu \Big ( {2 kT_e \over M_{\rm Fe} c^2}\Big )^{1/2} $,
where $\nu$ is the line frequency, $k$ is the Boltzmann constant, $T_e$
is the electron temperature, $M_{\rm Fe}$ is the mass of the iron ion, and
$c$ the speed of light,  
this corresponds to a few eV for the Fe K$_{\alpha}$ line. An X-ray 
spectrometer with adequate resolution could measure directly the 
attenuation in the line core.}. 
An alternative method to estimate the attenuation due to resonant 
scattering is to measure the Fe K$_{\alpha}$ line
and the Fe K$_{\beta}$ line. Since the 
 Fe K$_{\beta}$ line is expected to have an optical depth smaller than
1, the ratio of the Fe K$_{\beta}$ over Fe K$_{\alpha}$ equivalent widths
can lead to an estimate of the attenuation of the Fe K$_{\alpha}$ 
equivalent width and therefore of the real iron abundance.
The measurement of the Fe K$_{\beta}$ over Fe K$_{\alpha}$ line ratio 
relies critically on the measurement of the weaker of the 2 lines i.e.
the Fe K$_{\beta}$ line. The Medium Energy Concentrator/Spectrometer 
(MECS) 
on board Beppo-SAX is the first imaging X-ray experiment to have 
both sufficiently large effective areas at 8 keV to measure the line 
and sufficiently good spatial resolution in the 6-9 keV band to perform 
the measurement in the core of a nearby cluster such as Perseus. 
In this paper we present the measurement of the Fe K$_{\beta}$ over 
Fe K$_{\alpha}$ line ratio for the Perseus cluster.

Measurements of the Fe K$_{\beta}$ over Fe K$_{\alpha}$ ratio
have been performed in the past with various experiments.
Mitchell \& Mushotzky (1980), using HEAO1 A2 data, measured a ratio 
larger than would have been expected for an optically thin single 
temperature thermal plasma in the Centaurus cluster. 
Okumura et al. (1988), using TENMA data, found anomalous
line ratios in the Coma and Ophiuchus clusters. 
An observation of Perseus, presented in the same paper, lacked
the necessary S/N to derive a significant measure of the ratio.
Akimoto et al. (1997) have used ASCA data to analyze the
Fe K$_{\beta}$ over Fe K$_{\alpha}$ line ratios for a few nearby cluster,
namely: Abell 664, Virgo, Perseus, Abell 496 and Abell 3266.
They find that the Fe K$_{\beta}$ over Fe K$_{\alpha}$ ratios measured 
from their data are, for all these objects, in excess of what
is expected in the case optically thin thermal emission.
Unfortunately the above authors do not quote errors for their
measurements, thus making it difficult to assess the significance
of their results.  
A detailed modeling of the resonant scattering effect, 
performed through montecarlo simulations (Tawara et al. 1997) yields
predictions which are, according to Akimoto and coworkers, in 
disagreement with the measurements they have made on the clusters listed
above. Again the lack of errors on the data makes it difficult to
to understand how severe the disagreement is.

The outline of this letter is as follows.
In section 2 we give some information on the Beppo-SAX observation 
of Perseus and on the data preparation, a more complete presentation
will be given elsewhere (Molendi et al. in prep.). 
In section 3 we present the results of the analysis of the Fe K$_{\beta}$ 
and Fe K$_{\alpha}$ lines measured in a spectrum extracted from
a circular region, with radius 6.4 arcmin (corresponding to $\sim$ 200 kpc)
centered on the emission peak. We compare the observed line ratio
with that expected from an optically thin plasma and discuss a number 
of possible explanations for the difference between the two ratios.
In section 4 we measure the  Fe K$_{\beta}$ over  Fe K$_{\alpha}$ line
ratio for 5 concentric annuli centered on the emission peak. We argue that
the observed decrease in the Fe K$_{\beta}$ over  Fe K$_{\alpha}$ line 
ratio strongly favors the interpretation involving resonant scattering.
In section 5 we discuss the implications of the attenuation of the
Fe K$_{\alpha}$ line on the iron abundances in Perseus
and similar clusters.
In section 6 we summarize the main results of the paper.
Throughout the paper we assume a Hubble constant of
$H_o=50$ km s$^{-1}$Mpc$^{-1}$ and a redshift of $z=0.018$
for NGC 1275. Under the above assumptions an angular distance of 1 arcmin 
corresponds to 32 kpc. 
All spectral fits have been performed using XSPEC Ver. 9.01.
 Quoted confidence 
intervals are 68\% for 1 interesting parameter (i.e. $\Delta \chi^2 =1$), 
unless otherwise stated.

\section {Observation and Data Preparation}
The central region of the Perseus cluster was observed by the Beppo-SAX 
satellite (Boella et al. 97a) 
between the 20$^{th}$ and the 21$^{st}$ of September 1996
during the Science Verification Phase (SVP).
The observation was rather long with 
a total effective exposure of 89 ks for the Medium Energy Concentrator 
Spectrometer (MECS) 
(Boella et al. 97b). 
MECS data preparation and linearization was performed using 
the {\sc Saxdas v.1.0} package under {\sc Ftools} environment.

\section {Spectral Lines Analysis}


We have accumulated a  MECS spectrum from a circular region centered
on the emission peak using an extraction radius of 6.4 arcmin, corresponding
to $\sim$ 200 kpc.
We have modeled the spectrum in the 3-10 keV band using a bremsstrahlung for
the continuum and two Gaussian lines, one for the 6.8 keV iron complex,
the other for the 8 keV iron-nickel complex.
Data at energies 
below 3 keV has been excluded because significantly contaminated by the 
cooling flow (e.g. Fabian et al. 94).  
We find a temperature of 
$4.9\pm 0.1$ keV for the continuum component, an energy E$_1=6.76\pm 0.01$ 
keV and a width $\sigma_1 = 0.04\pm 0.04$ keV for the lower energy line 
and an energy E$_2=8.06\pm 0.04$ keV and a width $\sigma_2 =0.30\pm0.05$ 
keV for the higher energy line, where both E$_1$ and E$_2$ are given in
the source rest-frame.   
The 8 keV feature is clearly broad and most likely 
due to a blend of different lines from highly ionized 
iron (Fe XXV and Fe XXVI) and nickel (Ni XXVII). 

Fits with thermal emission codes such as MEKA, MEKAL
or Raymond \& Smith do not reproduce satisfactorily the 8 keV feature,
all these models underestimate significantly the intensity of the 
emission feature. In Figure 1 we present a fit of the MECS spectrum with a
MEKAL model, in the top panel we show the data and the folded 
model, in the bottom panel we  show the the residuals in the form 
of a ratio of the data over the model. A broad 
excess centered around 8 keV is clearly visible in the bottom panel.  
To better understand this discrepancy 
we have simulated thermal spectra at different temperatures using the
MEKAL, MEKA and  Raymond \& Smith codes and the MECS response matrix. 
We have then fitted the simulated data with a bremsstrahlung component 
for the continuum and 2 Gaussian lines, one for the 6.8 keV complex, 
the other for the 8 keV complex. Finally we computed the ratio of the 
flux in the 8 keV complex over the flux in the 6.8 keV complex. This 
ratio varies between 0.11 and 0.13 for temperatures in the range 4 keV 
to 8 keV. The ratio obtained from the real data, which is computed 
using the same spectral model applied to the simulated data (i.e.  
bremsstrahlung component for the continuum and 2 
Gaussian lines)  is $0.20\pm 0.02$. 

Since the MECS mirrors are made of gold coated nickel, and nickel 
fluorescence K$_{\alpha}$ lines are found $\sim$ 8 keV
we have performed a number of checks to verify that the emission feature
we see at 8 keV is not contaminated by a contribution of instrumental 
nature.
We have verified that the spectra of extremely bright galactic sources
such as Cyg X-1 and Crab show no evidence of such a feature. 
We have also verified that the spectrum of the instrumental
background, which in any case has an intensity $\simlt$ 1\% of the source
at 8 keV,  does not show any such feature.
Finally, in order to rule out possible Ni fluorescence from
the mirrors backside, we have analyzed the spectrum of the
galactic source GX 5-1 observed at an off-axis angle of 40 arcminutes.
Under such an observational configuration the fraction of photons that 
after having scattered once on the gold coated side of the mirrors, 
scatter a second time on the back side, made of nickel, 
and produce Ni fluorescence photons, is maximized. Even for this extreme
case we find no evidence of an emission feature $\sim$ 8 keV in the 
observed spectrum. 
For all these reasons we  conclude that the 8 keV feature is not
contaminated by instrumental contributions.

We have performed an independent measurement of the 8 keV over 6.8 keV 
line flux ratio using data collected on August 6$^{th}$ 1993 with the 
GIS detectors on board the ASCA satellite.
We accumulated GIS2 and GIS3 spectra from a circular region with the same 
radius (i.e. 6.4 arcmin) used for the MECS data. By fitting the 2 spectra
simultaneously in the energy range 3.5--10 keV,  with a bremsstrahlung 
component for the continuum and 2 Gaussian lines we find a line flux ratio
of $0.19\pm 0.03$. 
We note that the GIS measurement is in agreement with the MECS 
measurement.
We also note that the the difference between the 
ratio measured with the MECS and the expected one is statistically more 
significant than the difference between the ratio measured with the GIS 
and the expected one. The main reason for this is that the MECS 
observation (89 ks) is considerably longer than the GIS observation 
(17 ks). 

There are a number of possible explanations for the discrepancy between 
the observed line flux ratio and the expected one.
One possibility is that nickel is substantially over-abundant with 
respect 
to iron. If we fit the MECS spectrum in the energy range 3--10 keV with a
thermal emission model (MEKAL) imposing that the Ni abundance be the same
as the Fe abundance, we find a best fitting value of 
0.42$\pm {0.01}$ solar.  
If we fit the spectrum with a thermal model with free nickel 
abundances (VMEKAL) we find a significant improvement 
in the fit, ($>99$\% significance with an F-test) 
and a best fitting value of the nickel abundances of 
0.88$^{+0.22}_{-0.09}$ solar, the iron abundances remain  
unvaried. 
However, there are two\footnote
{Akimoto et al. (1997), who have investigated the 8.0 keV over 6.8 
keV line ratio in Perseus using ASCA GIS data, present a further
argument against the Ni overabundance. They remark that the Ni 
abundance required to explain the excess 8.0 keV emission 
is not consistent with the Ni abundance estimated from the L-shell Ni 
emission they measure $\sim$ 1 keV. The MECS data implies a Ni 
overabundance which is considerably smaller than the one estimated
in Akimoto et al. (1997), and is not inconsistent with the Ni 
abundance estimated by Akimoto et al. (1997) from the L-shell emission} 
points which make this interpretation difficult to support.
Firstly, the extreme Ni abundance is hard to reconcile with the Fe
abundances $\sim$ 0.4 solar (see Arnett 1995).
Secondly, although the inclusion of the line does improve the fit 
significantly, visual inspection of the residuals shows that 
the data is still in excess with respect to the model
around 8.4 keV, where the dominant contribution to the emission complex 
is expected to come from Fe XXV and Fe XXVI.
  
Another possibility is that the gas responsible for the
emission of the Fe 6.7-6.9 keV complex is not the same gas 
responsible for the emission of the 8 keV complex, indeed since the
emission around 6.8 keV is produced, on average, by iron that is less
ionized than that responsible for the emission at 8 keV, a particularly 
intense 8 keV component could be explained by the presence of a 
hot thermal component. Such a component would be characterized
by an  8 keV over 6.8 keV line flux ratio substantially larger than 
that expected from a component with a temperature of $\sim 5$ keV.  
We have tested this possibility by comparing
the ratio of the flux in the 8 keV complex over the flux in the 
6.8 keV complex expected from thermal emission codes such as   
MEKA, MEKAL and Raymond \& Smith to the observed ratio. The line ratios
expected from the thermal emission codes have been computed by 
performing fits on simulated spectra as already detailed in this section.
We find that even for extremely high temperature 
(i.e. $\sim$ 14 keV) all codes predict ratios smaller than 0.14 
while the observed ratio is 0.20$\pm$ 0.02. This result forces us 
to reject the two temperature interpretation, indeed if line emission
resulted from a mixture of gas at different temperatures, a significant
fraction of the gas would have to be at extremely high temperatures 
producing also continuum emission which is not observed. 

Another line of interpretation involves the ionization equilibrium 
of the gas. Emission codes assume that the gas is in collisional 
ionization equilibrium, with both ions and electron populations following 
a Maxwell-Boltzmann distribution, this may not be completely  true in the
central region of Perseus. Indeed NGC 1275 is known to harbor
an AGN which is extremely active at radio wavelengths (Nesterov et al. 
1995). Moreover, B\"ohringer et al. (1993)  by comparing radio  and 
X-ray high resolution images have found convincing evidence  of 
interaction
between the relativistic particles in the radio lobes of NGC 1275 and the 
intra-cluster plasma. 
One piece of evidence that argues against this possible interpretation
comes from the analysis of other clusters: 
Abell 644, Virgo, Abell 496 and Abell 3266 (Akimoto et al. 1997) all show 
8 keV over 6.8 keV line flux ratios 
in excess of the expected value. Thus an interpretation based on the unique 
characteristics of the central object in Perseus does not appear to be very 
attractive.

Another possibility is that the gas is indeed in thermal ionization 
equilibrium but the emission codes are wrong. Evidence of discrepancies 
between  the observed ratio of Fe XXVI over Fe XXV with respect to 
the expected one has been found in solar flares (Tanaka 1986). Although,
as mentioned in Arnaud \& Raymond (1992), one cannot exclude that the 
X-ray emitting gas in solar flares is not in collisional equilibrium.  

The last mechanism we consider to explain the observed anomalous
ratio is the photon redistribution due to resonant scattering.
As pointed out in G86,
the inter-galactic medium in clusters,
while optically thin to the continuum and many lines, may be optically
thick in the center of some lines. Let us consider first the 6.8 keV
complex. It is dominated by H-- and especially He--like recombination
lines. At these temperatures the strongest line in the complex is
the 6.7 keV He--like resonant line, which accounts for more than 50\% 
of the total emission. Its central optical thickness along the whole 
cluster is, for Perseus, about 1.5-2, adopting the simplifying 
assumption of constant temperature, kT$\sim$ 6 keV, and abundances, 
$\sim 0.5$ solar, and a $\beta$=2/3 model for the density, with central 
density of 5$\times$10$^{-3}$ cm$^{-3}$; see G86 and Sarazin (1988), 
for the relevant formulae. (Note that this is likely to be and 
underestimate, due to the neglect of the cooling flow). The two 
most important H--like lines, localized at 6.97 keV, which are both 
resonant, as well as the inter-combination 6.67 keV
He--like line, accounting all together for most of the remaining emission 
of the complex, have central thicknesses an order of magnitude
lower. The 6.65 keV He--like forbidden line has a negligible depth, 
but it accounts for no more than the 15\% or so of the total emission.
On the other hand, the 8 keV complex is dominated by the Ni He--like,
K$_\alpha$ lines, 7.8 keV, and
the Fe He--like K$\beta$ lines, 7.9 and 8.2 keV, which have optical 
depths about 20 and 7 times smaller than the 6.7 keV He--like line. 
It is therefore clear 
that optical depth effects are considerably more relevant for the 6.8 keV 
complex rather than for the 8 keV one.

After the absorption, all photons are
re--emitted, as for these ions Auger de--excitation cannot work. 
The process is then purely diffusive (resonant scattering) and has the 
effect of redistributing the line photons over the cluster.
Assuming spherical symmetry, the effect is to decrease the observed
line emissivity in the central region, where the
line--of--sight column density is greater, balanced by an excess
emissivity in the outer regions. As this redistribution effect is greater
for larger optical depths, we expect a gradient in the ratio of the two
line complexes. We have therefore performed a spatially resolved analysis,
which is described in the next section.

\section{Spatially Resolved Spectral Analysis} 


We have accumulated spectra from 5 concentric annuli centered
on the emission peak and with bounding radii of 0-2 arcmin, 2-4 arcmin,
4-6 arcmin, 6-8 arcmin and 8-10 arcmin respectively.  We have fitted each 
spectrum in the 3.5-9.5 keV range.
We have modeled the spectra with a bremsstrahlung component for
the continuum and 2 Gaussian lines, one for the 6.8 keV complex,
the other for the 8 keV complex. The ratio of the flux in the lines
varies with the radius (see Fig.2), a $\chi^2$ test rejects a 
constant line ratio at the 98\% confidence level. 
The solid line in Figure 2 is the ratio expected in the case of 
optically thin thermal emission for temperatures $\sim$ 5 keV.
The dotted line represents the expected ratio
when the 8 keV line complex flux in excess of  what is expected from 
an optically thin thermal plasma is distributed as
if it came from a point source coincident with the X-ray emission peak.

Akimoto et al. (1997) have performed a similar analysis of the line
ratio in Perseus using ASCA GIS data. They find a radial gradient 
which appears to be different 
from ours. It is rather difficult to make a more detailed comparison, 
mainly because in their Figure 1: 1) the radii are reported in units 
of the core radius, but the value of the core radius is not given; 2)  
errors are not shown so we cannot assess how significant the difference
actually is. 
The same authors present a radial profile of the expected line ratio
based on numerical simulations which include the effects of resonant
scattering. This profile appears to be in disagreement with our 
measurements as well as theirs. Even though the authors do not give
many details on the parameters chosen for the simulation, from 
the line ratio predicted in the case of optically thin emission, also
shown in their Figure 1, it would seem that they assume a strong
temperature gradient within the core of Perseus, possibly to mimic
the cooling flow. An analysis of the MECS data shows that, excluding
energies below 3 keV where the cooling flow is prominent, the radial
temperature gradient in the innermost 200 kpc is rather modest,  
$\Delta T \sim 1.5$ keV. If this is indeed the case then the 
decrement in the Fe K$_{\alpha}$ line complex intensity may be 
considerably larger than that estimated through simulations
in Akimoto et al. (1997). 


We note that explanations of the anomalous line flux ratio  
involving the ionization equilibrium cannot easily 
explain the results of the spatially resolved analysis.
Indeed if the ionization equilibrium is not thermal because 
of the active nucleus in  NGC 1275, we would expect the ratio to be
anomalous only in the innermost spatial bin (the radio
lobes extend out to less than 1 arcmin form the nucleus). 
The data shows that this is not the case. Moreover if the 
anomalous ratio is the result of an error in the thermal 
emission codes then, of course, the ratio should be the same 
throughout the cluster.

The interpretation in terms of photon redistribution by resonant 
scattering appears to be the only viable one among those considered. 
It is rather difficult to check it quantitatively, as one should consider
in detail the abundance and temperature gradients, as well as deviations
from a simple King profile for the density due to the cooling flow.
The treatment of this problem in real cluster conditions then requires a
detailed radiative transfer solution, and is deferred to future works.
We here limit ourselves to noting that, according to G86 calculations,
the observed factor--of--2 relative depletion of the 6.8 keV complex
in the cluster center is consistent with the order--of--magnitude 
difference in the optical depth of the two complexes.
 
\section{Abundances} 


From the analysis of the spatially integrated and resolved 
spectra of the core of Perseus we have found that
resonant scattering provides the only acceptable explanation 
for the anomalous line flux ratio.  
In this section we discuss the implications of this result.

If we accept that emission from the core of Perseus 
is optically thick at the Fe K$_{\alpha}$ line then
the the common assumption that the equivalent width of the line
can be used to directly measure the abundance of iron breaks down.
As already noted in G86,
iron abundance measures derived by fitting spectra 
with standard thermal emission codes, which assume emission
to be optically thin at all wavelengths, will significantly 
underestimate the true abundance.

We have computed an ``apparent'' abundance radial profile by fitting 
the spectra from 4 concentric annuli with bounding radii, 0-2 arcmin,
2-4 arcmin, 4-6 arcmin and 6-8 arcmin, with a MEKAL model in the 
energy range 3.5-7.7 keV (energies larger than 7.7 keV have been excluded
to avoid fitting the 8 keV complex). We find that, the measured 
Fe abundance is largest at the center with a value of 0.48$\pm 0.01$ 
solar, and decreases with 
increasing radius, reaching a value of 0.29$\pm 0.01$ solar at  
6--8 arcmin ($\sim$ 220 kpc) from the cluster center 
(see Fig. 3). 
   
Under the simplyfing assumption that the lines contributing to the 8 keV 
blend are totally unaffected by resonant scattering 
an estimate of the real iron Abundance profile can be obtained 
in two different ways: 
by dividing the values of the ``apparent'' abundances obtained through 
spectral fitting with the MEKAL  model by the ratio between the observed 
line flux ratio and the line flux ratio expected for optically thin
emission; or by measuring the abundances directly with the Fe
K$_{\beta}$ line, this can be done by re-fitting the observed spectra with 
the MEKAL model having first excluded the region of the 
spectrum, 6--7.5 keV, where the K$_{\alpha}$ line  contributes
significantly to the total emission.
Abundances estimated in either of these two fashions should be 
regarded as lower limits, indeed if some of the lines contributing to 
the 8 keV blend are attenuated by resonant scattering 
the real abundances will be even larger than the ones we compute. 
The two methods yield values of the abundances which are always in 
agreement, although the abundances obtained by fitting the K$_{\beta}$
line are 
somewhat smaller than those obtained by correcting the ``apparent'' 
abundances (see Fig. 3). The reason for this difference is likely due
to the difficulty of excluding completely the contributions of the
Fe K$_{\alpha}$ line to the observed spectrum when computing the
abundances with the Fe K$_{\beta}$ line.  
Both methods yield an abundance  profile characterized 
by a gradient which is considerably larger than the one observed in the
``apparent'' abundance profile.
The Fe abundance in the innermost circular region, with radius 2 arcmin,  
corresponding to $\sim$ 30 kpc,  is  $\sim$ 0.9 solar and consistent with 1.
The obvious implication is that a very large fraction of the gas in 
the core of Perseus has been processed in stars.  
It is interesting to note that the abundance gradient extends out to 
$\sim 200$ kpc from NGC 1275. 
This scale length is comparable to the one characterizing 
the cooling flow observed in the core of Perseus 
(e.g. Fabian et al 1981). The association of these two phenomena,
i.e. the abundance gradient and the cooling flow, may not be casual.
One could speculate that an iron rich gas outflow from NGC 1275
may have increased the gas density in the core of Perseus and 
consequently have triggered the cooling flow.

We do not expect iron lines emission from the AGN in NGC 1275 to
alter in a significant way our estimate of the abundance in the innermost
radial bin. 
Observations in the hard X-ray band with the PDS instrument on board 
BeppoSAX (Molendi et al. in preparation) indicate that the AGN should 
contribute less than 10\% to the continuum emission in the 6-9 keV band.
Moreover eventual contributions to the Fe K$_{\beta}$ line complex at 8 keV,
which we have used to estimate the iron abundance, 
should be negligible, since iron line emission in AGN is observed in
the 6-7 keV band and practically never around 8 keV.

The values of the gas density and temperature in the core 
of the Perseus cluster imply an optically thick Fe K$_{\alpha}$  
line (see G86).
Similar values of the gas density and temperature are measured in the core 
of many rich clusters, thus resonant scattering is likely to operate in many 
objects. The high Fe K$_{\beta }$ over Fe K$_{\alpha}$ flux ratio 
measured in the core of 
Abell 644, Virgo, Abell 496 and Abell 3266 (Akimoto et al. 1997)   
clearly supports this possibility.
If resonant scattering operates effectively in the core of many clusters 
than iron may be considerably more abundant, in these objects, than we have
so far believed. 

\section{Summary} 
The analysis of the MECS spectrum of the central region of Perseus
has shown that the ratio of the flux in the 8.0 keV line complex 
over the flux in the 6.8 keV line complex is about a factor 2 larger
than expected for an optically thin thermal plasma. Moreover this
ratio presents a radial gradient, in the sense that it decreases 
with increasing cluster radius.
We have considered various possible explanations for the  
behavior of the lines ratio: 1) instrumental effects; 2) a high nickel abundance; 
3) a second hot thermal component; 4) non thermal ionization 
equilibrium; 5) errors in the computation of the collisional ionization 
equilibrium; 6) resonant scattering of the Fe K$_{\alpha}$ line. 
We have argued that resonant scattering is, by far, the most likely 
explanation for the observed line ratio behavior.

We have discussed the implications of an optically thick 
Fe K$_{\alpha}$ line. We have found the Fe abundances in the 
core of Perseus to be significantly larger than previously believed,
the true abundances being extremely high ($ \simgt 0.9 $ solar in a 
circular region with a radius of $\sim 60$ kpc centered on the
emission peak). We have argued that similar 
underestimations of the Fe abundance may have also occurred for other rich 
clusters. 

\acknowledgments
We acknowledge support from the BeppoSAX Team. SM thanks 
Paolo Saracco and Mario Zannoni for useful discussions and 
Sandro Mereghetti for access to GX5-1 data prior to publication.

\clearpage


\clearpage

\figcaption
{MECS Spectrum of the core of the Perseus cluster, data is 
extracted from a circular region with a radius of 6.4 arcminutes
corresponding to $\sim$ 200 kpc.  In the top panel we show the 
data and the best fitting MEKAL model. In the bottom panel 
we show the residuals in the form of a ratio of the data over
the model. }

\figcaption
{Ratio of the 8 keV line complex flux over the 6.8 keV line complex flux vs.
apparent cluster radius.
The solid line represents the expected ratio for an optically thin plasma
at a temperature of 5 keV. The dotted line represents the expected ratio
when all the 8 keV line complex flux in excess of  what is expected from 
an optically thin thermal plasma is distributed as
if it came from a point source coincident with the X-ray emission peak.}

\figcaption
{Iron radial abundance profile. The ``apparent'' abundances,
measured using the Fe K$_{\alpha}$ line under the assumption of
optically thin emission, are indicated with the open circles.
The filled circles indicate the abundances obtained by correcting
the ``apparent'' abundances (see text for details). The open squares
indicate the abundances obtained using the Fe K$_{\beta}$ line.}

\clearpage

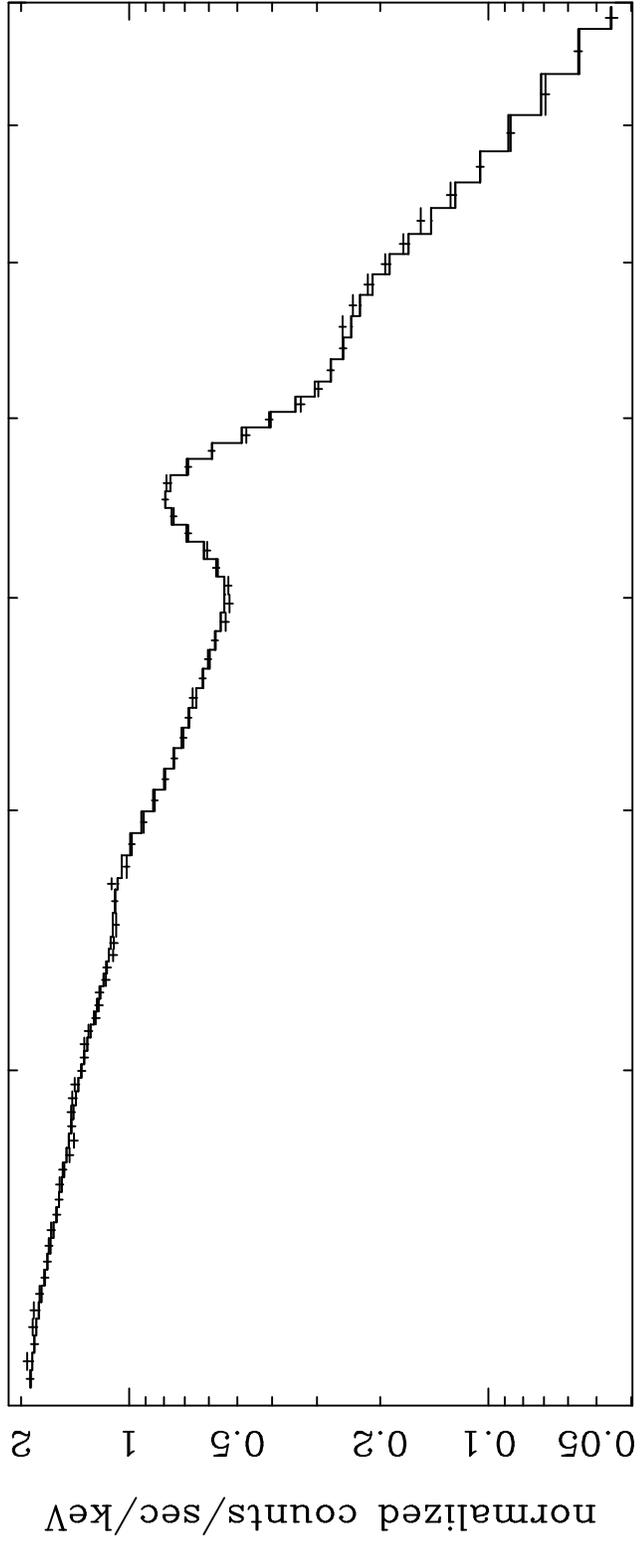
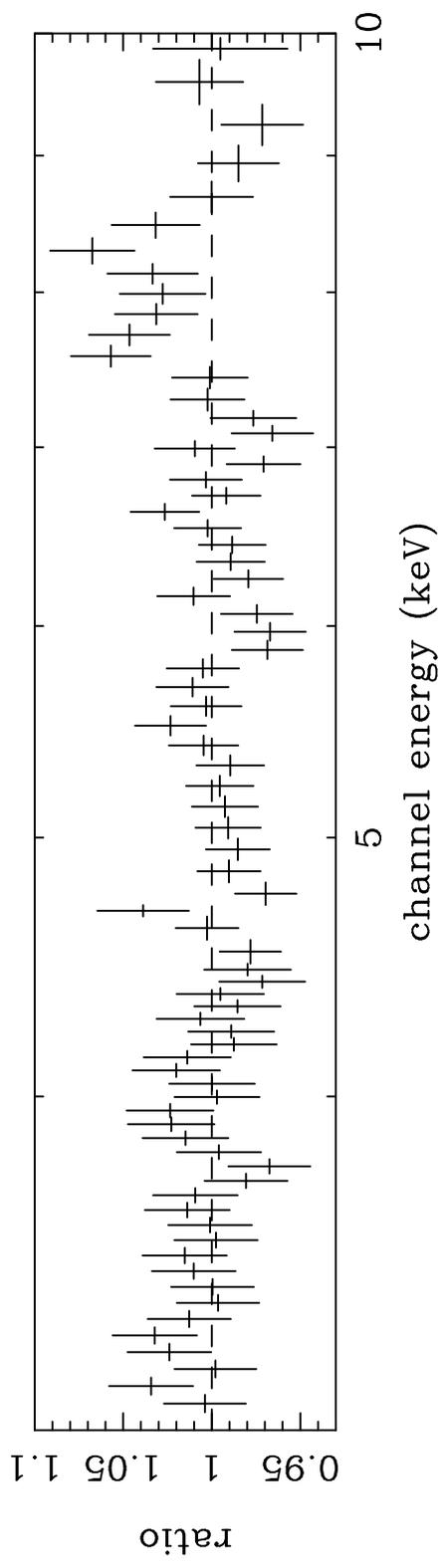
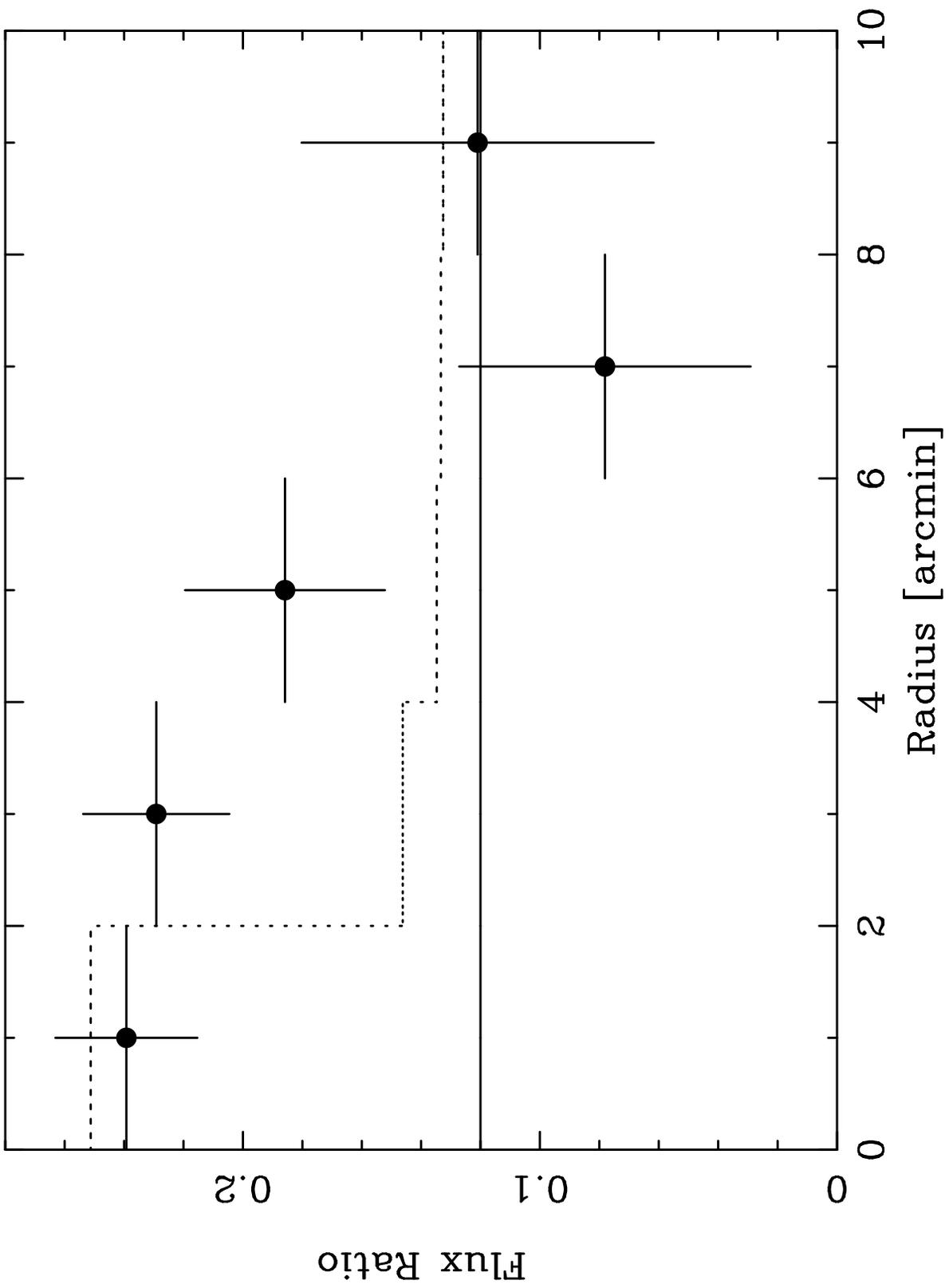
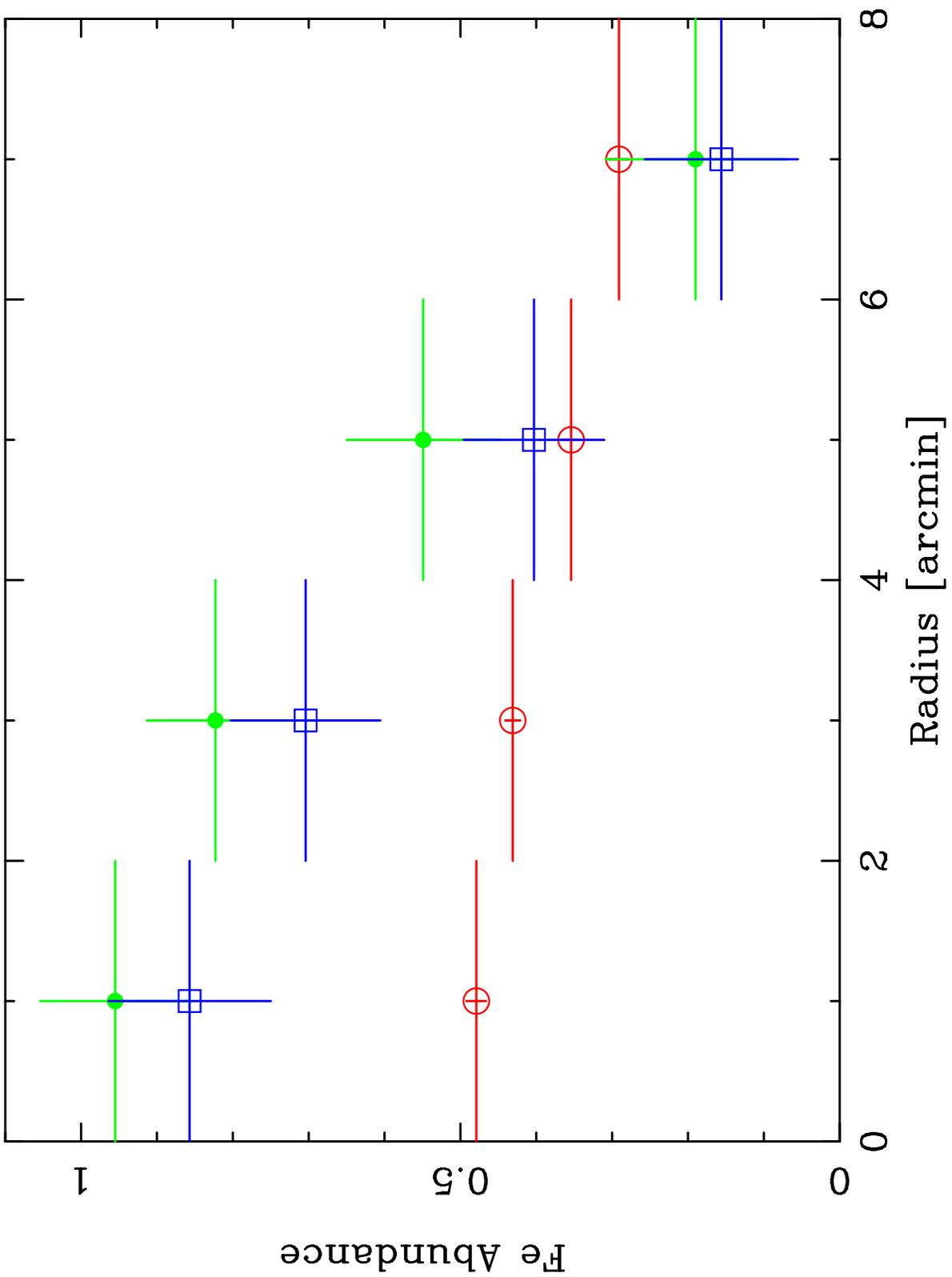
\end{document}